\documentstyle[aps]{revtex}

\newcommand{\be}{\begin{equation}}
\newcommand{\ee}{\end{equation}}
\newcommand{\ba}{\begin{eqnarray}}
\newcommand{\ea}{\end{eqnarray}}
\newcommand{\er}{\end{eqnarray}}
\newcommand{\br}{\begin{eqnarray}}
\newcommand{\dslash}{\partial\!\!\!/}
\newcommand{\aslash}{a\!\!\!/}
\newcommand{\bslash}{b\!\!\!/}

\begin{document}



\title{ Duality Equivalence Between Self-Dual And Topologically Massive Non-Abelian Models}
\author{A. Ilha and C. Wotzasek}

\address{Instituto de F\'{\i}sica, Universidade Federal do Rio de Janeiro,\\
         Caixa Postal 68528, 21945 Rio de Janeiro, RJ, Brazil.}
\date{\today}
\maketitle

\begin{abstract}

\noindent The non-abelian version of the self-dual model proposed by Townsend, Pilch and van Nieuwenhuizen presents some well known difficulties not found in the abelian case, such as well defined duality operation leading to self-duality and dual equivalence with the Yang-Mills-Chern-Simons theory, for the full range of the coupling constant.  These questions are tackled in this work using a distinct gauge lifting technique that is alternative to the master action approach first proposed by Deser and Jackiw. The master action, which has proved useful in exhibiting the dual equivalence between theories in diverse dimensions, runs into trouble when dealing with the non-abelian case apart from the weak coupling regime.
This new dualization technique on the other hand, is insensitive of the non-abelian character of the theory and generalize straightforwardly from the abelian case.
It also leads, in a simple manner, to the dual equivalence for the case of couplings with dynamical fermionic matter fields.
As an application, we discuss the consequences of this dual equivalence in the context of 3D non-abelian bosonization.
\end{abstract}
\newpage

\section{Introduction}

The bosonization technique that expresses a theory of interacting fermions in terms of free bosons provides a powerful non-perturbative tool for investigations in different areas of theoretical physics with practical applications \cite{2dbos}.
In two-dimensions these ideas have been extended in an interpolating representation of bosons and fermions which clearly reveals the dual equivalence character of these representations \cite{DNS}.
In spite of some difficulties, the bosonization program has been extended to higher dimensions \cite{luscher,marino}.

In particular the $2+1$ dimensional massive Thirring model (MTM) has been bosonized to a free vectorial theory in the leading order of the inverse mass expansion.
Using the well known equivalence between the self-dual \cite{TPvN} and the topologically massive models\cite{annals} proved by Deser and Jackiw \cite{DJ} through the master action approach \cite{suecos}, a correspondence has been established between the partition functions for the MTM and the Maxwell-Chern-Simons (MCS) theories \cite{FS}.  The situation for the case of fermions carrying non-abelian charges, however, is less understood due to a lack of equivalence between these vectorial models, which has only been established for the weak coupling regime \cite{BFMS}.  As critically observed in \cite{KLRvN} and \cite{BBG}, the use of master actions in this situation is ineffective for establishing dual equivalences.  In this paper we intend to fill up this gap.

We propose a new technique to perform duality mappings for vectorial models in any dimensions that is alternative to the master action approach.
It is based on the traditional idea of a local lifting of a global
symmetry and may be realized by an iterative embedding of Noether counter terms.
This technique was originally explored in the context of the
soldering formalism \cite{ABW,BW} and is exploited here since it seems to be the most appropriate technique for non-abelian generalization of the dual mapping concept.

Using the gauge embedding idea, we clearly show the dual equivalence between the non-abelian self-dual and the Yang-Mills-Chern-Simons models, extending the proof proposed by Deser and Jackiw in the abelian domain.
These results have consequences for the bosonization identities from the massive Thirring model
into the topologically massive model, which are considered here, and also allows for the extension of the fusion of
the self-dual massive modes \cite{BW} for the non-abelian case \cite{IW2}.
We also discuss the case where charged dynamical fermions are coupled to the vector bosons. For fermions carrying a global U(1) charge we reproduce the result of \cite{GMdS} but the result for the non-abelian generalization is new.

The technique of local gauge lifting is developed in section II through specific examples.
In section III we show how the gauge embedding idea solves the problem of non-abelian dual equivalence.
For completeness we first discuss the abelian case showing how the well known results are easily reproduced.
The case of dual equivalence between the SD and the MCS when dynamical fermion fields are coupled to the gauge fields is also considered, both in the abelian and in the non-abelian cases. The remaining sections are dedicated to explore this result in the non-abelian bosonization program and to present our conclusions.

\section{Noether Gauge Embedding Method}

Recently there has been a number of papers examining the existence of gauge invariances in systems with second class constraints \cite{vithee1}.
Basically this involves disclosing, using the language of constraints, hidden gauge symmetries in such systems.
This situation may be of usefulness since one can consider the non-invariant model as the gauge fixed version of a gauge theory.  By doing so it has sometimes been possible to obtain a deeper and more illuminating interpretation of these systems.
Such hidden symmetries may be revealed by a direct construction of a gauge invariant theory out of a non-invariant one \cite{neves}.
The former reverts to the latter under certain gauge fixing conditions.
The associate gauge theory is therefore to be considered as the embedded one. The advantage in having a gauge theory lies in the fact that the underlying gauge invariant theory allows us to establish a chain of equivalence among different models by choosing different gauge fixing conditions.

In this section we shall review a different technique to
achieve this goal: the iterative Noether gauging procedure.
For pedagogical reasons, we develop simple illustrations making use of scalar field theories living in a Minkowski space-time of dimension two.
This will allow us to discuss some subtle technical details of this method, regarding the connection between the implementing symmetries and the Noether currents, which are necessary for its application in the 2+1
dimensional self-dual model.

The important point to stress in this review of the iterative Noether procedure is its ability to implement specific symmetries leading to distinct models.  To avoid unnecessary complications let us consider the case of a free two dimensional scalar field theory,

\be
\label{QW05}
{\cal S}^{(0)}= \frac 12 \int\, d^2 x \;
\partial_\mu \varphi \: \partial^\mu \varphi\,,
\ee
and choose to gauge either the axial shift,

\be
\label{QW10}
(i) \;\;\;\;\; \varphi\to\varphi + \epsilon\,,
\ee
or the conformal symmetry,

\be
\label{QW20}
(ii) \;\;\;\;\; \varphi\to\varphi + \epsilon \; \partial_-\varphi\,, \ee
where $\partial_\pm = \frac 1{\sqrt 2}\left(\partial_0 \pm \partial_1\right)$ and $\epsilon$ is a global parameter.
In the first case it is simple to identify the Noether current as,

\be
\label{QW40}
J_\mu = \partial_\mu\varphi\,,
\ee
which comes from the variation of the scalar field action

\be
\label{QW30}
\delta {\cal S}^{(0)} = \int\, d^2 x \; J_\mu \:\partial^\mu\epsilon\,,
\ee
which is non-vanishing if $\epsilon$ is lifted to its local version. To compensate for this non-vanishing
result we introduce a counter-term together with an ancillary gauge field $B_\mu$ (also called as compensatory field)
as,

\be
\label{QW50}
{\cal S}^{(1)} = {\cal S}^{(0)} - \int\, d^2 x\; J_\mu B^\mu\,,
\ee
such that its variation reads,

\be
\label{QW60}
\delta {\cal S}^{(1)} = - \int\, d^2 x\; B^\mu \delta J_\mu\,,
\ee
which is achieved if $B_\mu$ transforms as a vector field simultaneously with (\ref{QW10}).  Introducing an extra
counter-term as,

\be
\label{QW70}
{\cal S}^{(2)} = {\cal S}^{(1)} + \frac 12 \int\, d^2 x\: B_\mu B^\mu\,,
\ee
that automatically compensates for (\ref{QW60}), will render ${\cal S}^{(2)}$ gauge invariant, ending the iterative chain.

To implement, on the other hand, the chiral diffeomorphism (\ref{QW20}) at a local level we proceed
analogously but starting with the scalar action written in terms of light-cone variables as,

\be
\label{QW80}
{\cal S}^{(0)}= \int\, d^2 x \:\partial_+ \varphi \;\partial_- \varphi \, .
\ee
We then obtain,

\be
\label{QW90}
\delta {\cal S}^{(0)} = \int\, d^2 x \:T(\varphi)\: \partial_+\epsilon\,,
\ee
after discarding some total derivatives, with

\be
\label{QW95}
T(\varphi)=(\partial_-\varphi)^2\,,
\ee
being a component of the energy-momentum tensor. The first iterated action,
that compensates for this non-null result requires a new gauge field $\lambda$ as,

\be
\label{QW100}
{\cal S}^{(1)} = {\cal S}^{(0)} - \int\, d^2 x\: \lambda \: T(\varphi)\,.
\ee
After some algebra one finds

\be
\label{QW110}
\delta {\cal S}^{(1)} =  \int\, d^2 x \:\left[\partial_+\epsilon - \delta \lambda - 2 \lambda\: \partial_-\epsilon +
\partial_-\left(\lambda\:\epsilon\right)\right]\:T(\varphi)\,,
\ee
which vanishes identically if $\lambda$ is chosen to transform as a component of
the metric tensor under the chiral diffeomorphism (\ref{QW20}),

\be
\label{QW120}
\delta \lambda = \partial_+\epsilon  - \lambda\partial_-\epsilon +
\epsilon\partial_-\lambda\,.
\ee
We mention that the gauge invariant action (\ref{QW100}) has been first proposed by Siegel \cite{WS} to represent the dynamics of two-dimensional self-dual scalar fields.

It is interesting to observe that, in the canonical approach, the Noether currents (\ref{QW40}) and (\ref{QW95})
become the generators of the transformations (\ref{QW10}) and (\ref{QW20}) respectively.  In Dirac's nomenclature
of constrained systems \cite{PAMD}, these currents are first-class gauge generator constraints  derived from the gauge converted model.  It is important to mention that the original models may be reobtained from these gauge invariant ones
by choosing the so called unitary gauge but different gauge choices are now at our disposal.

\section{Dual Equivalence of SD and MCS Theories}

In this section we discuss the application of the gauge invariant embedding to show the dual equivalence between the SD model with the MCS theory both abelian and non-abelian including the coupling with charged dynamical matter fields.  As mentioned in the introduction this has the advantage of possessing a straightforward extension to the non-abelian case for all values of the coupling constant.

Let us recall that the essential properties manifested by the three dimensional self-dual theory such as parity breaking and anomalous spin,
are basically connected to the presence of the topological and gauge invariant Chern-Simons term. The abelian self-dual model for vector fields was first introduced by Townsend, Pilch and van
Nieuwenhuizen \cite{TPvN} through the following action,
\be
\label{180}
{\cal S}_{\chi}[f]= \int\, d^3x\:\Bigg(\frac {\chi }{2m}\,
\epsilon_{\mu\nu\lambda}\,f^\mu\,\partial^\nu f^\lambda -
\frac{1}{2}\, f_\mu f^\mu \Bigg)\,,
\ee
where the signature of the topological terms is dictated by $\chi = \pm 1$
and the mass parameter $m$ is inserted for dimensional reasons.
Here the Lorentz indices are represented by greek letters taking their usual values
as $\mu , \nu , \lambda = 0,1,2$.
The gauge invariant combination of a Chern-Simons term with a Maxwell action
\be
\label{370}
{\cal S}^{(MCS)}= \int\, d^3x\:\Bigg(\frac{1}{4m^{2}} f^{\mu\nu}f_{\mu\nu} -
\frac{\chi}{2m}\,\epsilon^{\mu\nu\lambda}\,f_{\mu}\,
\partial_{\nu}f_{\lambda}\Bigg)\,,
\ee
is the topologically massive theory, which is known to be equivalent \cite{DJ} to the self-dual model (\ref{180}).
$f_{\mu\nu}$ is the usual Maxwell field strength,

\be
\label{285}
f_{\mu\nu} \equiv \partial_{\mu}f_{\nu} - \partial_{\nu}f_{\mu}\,.
\ee

The non-abelian version of the
vector self-dual model (\ref{180}), which is our main concern in this work, is given by
\be
{\cal S}_\chi = \int\,d^{3}x\, tr\left[\,-\frac{1}{2}\,{ F}_{\mu}\,{ F}^{\mu} +
\frac{\chi}{4m}\,\epsilon^{\mu\nu\lambda}\left({ F}_{\mu\nu}\,
{ F}_{\lambda}
- \frac{2}{3}\,{ F}_{\mu}\,{ F}_{\nu}\,{ F}_{\lambda}
\right)\,\right]\,,
\label{380}
\ee
where ${F}_{\mu} = F_{\mu}^{a}{t}^{a}$, is a vector field taking values in the Lie algebra of a symmetry group $G$ and ${ t}^{a}$ are the matrices
representing the underlying nonabelian gauge group with $a= 1,\ldots , \mbox{dim}\:G$.
The field-strength tensor is given by
\be
{ F}_{\mu\nu} = \partial_{\mu}{ F}_{\nu} - \partial_{\nu}{ F}_{\mu} +
\left[{ F}_{\mu}\,,\,{ F}_{\nu}\right]\,,
\label{390}
\ee
and the covariant derivative is ${ D}_{\mu} = \partial_{\mu} + [F_{\mu}\,,\, ]$.

Using the master action approach, the action (\ref{380}) has been shown to be equivalent to the gauge invariant Yang-Mills-Chern-Simons (YMCS) theory
\be
{\cal S}^{(YMCS)} = \int\,d^{3}x\,tr\left[\,
\frac{1}{4m^{2}}\,{ F}^{\mu\nu}\,{ F}_{\mu\nu} +
\frac{\chi}{4m}\,\epsilon^{\mu\nu\lambda}\left({ F}_{\mu\nu}\,
{ F}_{\lambda}
- \frac{2}{3}\,{ F}_{\mu}\,{ F}_{\nu}\,{ F}_{\lambda}\right)
\,\right]\,,
\label{510}
\ee
only in the weak coupling limit $g\to 0$ so that the Yang-Mills term effectively vanishes
\footnote{Here we are using the bosonization nomenclature that relates the Thirring model coupling constant $g^2$ with the inverse mass of the vector model; see discussion after Eq.(\ref{NB100})}.
To study the dual equivalence of (\ref{380}) and (\ref{510}) for all coupling regimes and the consequences over the bosonization program is main contribution of this work.

Next we analyze the dualization procedure in the massive
spin one self-dual theories using the Noether gauging procedure.
To begin with, it is useful to clarify the meaning of the self duality inherent in the
action (\ref{180}). The equation of motion in the absence of sources is given by,
\be
\label{190}
f_\mu =\frac{\chi}{m}\,
\epsilon_{\mu\nu\lambda}\,\partial^\nu f^\lambda\,,
\ee
from which the following relations may be easily verified,
\br
\label{200}
\partial_\mu f^\mu &=& 0\,,\nonumber\\
\left(\Box + m^2\right)f_\mu &=& 0\,.
\er
From (\ref{190}) and (\ref{200}) we see that there is only one massive excitation whose value is $m$.
A field dual to $f_\mu$ is defined as,
\be
\label{210}
{}^{\star}f_\mu = {1\over m}\, \epsilon_{\mu\nu\lambda}\,
                  \partial^\nu f^\lambda\,.
\ee
Repeating the dual operation, we find,
\be
\label{220}
{}^{\star}\left({}^{\star}f_{\mu}\right)=
{1\over m}\, \epsilon_{\mu\nu\lambda}\,\partial^\nu\,
{}^{\star}f^\lambda = f_\mu\,,
\ee
obtained by exploiting (\ref{200}), thereby validating the definition of
the dual field.  Combining these results with (\ref{190}), we conclude that,
\be
\label{230}
f_\mu= \chi\,\, {}^{\star}f_\mu\,.
\ee
Hence, depending on the signature of $\chi$, the theory will correspond to a
self-dual or an anti self-dual model.

To prove the exact equivalence between the self-dual model and the Maxwell
Chern-Simons theory, we start with the zeroth-iterated action (\ref{180})
which is non-invariant by gauge transformations of the basic vector field $f_{\mu}$.
To construct from it an abelian gauge model,  we have to consider the
gauging of the following symmetry,
\be
\label{260}
\delta f_\mu = \partial_{\mu}\xi\,,
\ee
where $\xi$ is an infinitesimal local parameter.
Under such transformations, the action (\ref{180}) change as,
\be
\label{270}
\delta {\cal S}_{\chi} = \int\,d^{3}x\,J^{\mu}(f)\,\partial_{\mu}\xi\,,
\ee
where the Noether currents are defined by,
\be
\label{280}
J^{\mu}(f) \equiv - f_{\mu} + \frac{\chi}{2m}\,\epsilon_{\mu\nu\lambda}\,
            f^{\nu\lambda}\,,
\ee
and $f_{\mu\nu}$ is the usual Maxwell field strength (\ref{285}).
It is worthwhile to mention that any other variation of the fields
(like $\delta{f_\mu}=\xi_\mu$) is inappropriate because changes in
the two terms of the Lagrangians cannot be combined to give a single
structure like (\ref{280}). We now introduce an ancillary vector
field $B_{\mu}$ coupled with the Noether currents and transforming in the
usual way,
\be
\label{290}
\delta B_{\mu}= \partial_{\mu} \xi\,.
\ee
Note that in the abelian case, $\delta f_{\mu} = \delta B_{\mu}$.
Then it is easy to see that the modified action,
\be
\label{300}
{\cal S}^{(1)} = {\cal S}_{\chi} - \int\,d^{3}x\,\,J_{\mu}\,B^{\mu}\,,
\ee
transform as,
\be
\label{310}
\delta {\cal S}^{(1)} = - \, \int\,d^{3}x\,\delta J_{\mu}\, B^{\mu}\,.
\ee
The final modification consists in adding a term to ensure gauge invariance.
This is achieved by,

\be
\label{330}
{\cal S}^{(2)} = {\cal S}_{\chi} - \int\,d^{3}x\,\left(J^{\mu}\,B_{\mu} + \frac{1}{2}\,B^{\mu}\,B_{\mu}\right)\,,
\ee
which is invariant under the combined gauge transformations (\ref{260}) and (\ref{290}).
The gauging of the U(1) symmetry is complete. To return to a description
in terms of the original variables, the auxiliary vector field is
eliminated from (\ref{330}) by using the equation of motion,
\be
\label{340}
B_{\mu} = - J_{\mu}.
\ee
Note that taking variations on both sides of this equation and using the gauge invariance of the Chern-Simons form we obtain consistency with the condition $\delta f_{\mu} = \delta B_{\mu}$.
It is now crucial to note that, by using the explicit structures for the
currents, the above action (\ref{330})  forms a gauge invariant
combination expressed by the action (\ref{370})
which is the Maxwell Chern-Simons theory.
Our goal has been achieved. The iterative Noether dualization procedure has precisely incorporated
the abelian gauge symmetries in the self-dual model to yield the gauge
invariant Maxwell Chern-Simons theory.

The free case considered above can also be extended to couplings with
external, field-independent sources. To illustrate this point, we consider
a coupling between dynamical U(1) charged fermions and self-dual vector bosons \cite{GMdS}. In the abelian
case, this model is written as
\be
{\cal S}[f,\psi] = {\cal S}_{\chi}[f] + \int\,d^{3}x\,\left(
- e\,f^{\mu}\,J_{\mu} + \bar{\psi}\left(i\,\dslash - M\right)\psi\right)\,,
\ee
where $J_{\mu} = \bar{\psi}\,\gamma_{\mu}\,\psi$
and $M$ is the fermionic mass. ${\cal S}_{\chi}$ is the self-dual
action (\ref{180}). The Noether current associated with this action
is
\be
K_{\mu} = -f_{\mu} + \frac{\chi}{m}\,\epsilon^{\mu\nu\lambda}\,
\partial_{\nu}f_{\lambda} - e\,J_{\mu}\,.
\ee
By introducing an ancillary field $B_{\mu}$, we can construct a gauge
invariant theory which reads
\be
{\cal S}_{B} = {\cal S}[f,\psi] - \int\,d^{3}x\,\left(
K^{\mu}\,B_{\mu} +\frac{1}{2}\,B^{\mu}\,B_{\mu}\right)\,.
\ee
After the elimination of $B_{\mu}$ through its equations of motion, we
get our final theory,
\be
\label{9785}
{\cal S} = {\cal S}^{(MCS)} + \int\,d^{3}x\,\left(e^{2}\,J^{\mu}\,J_{\mu}
 + e\,\chi f_\mu G^\mu  + {\cal L}_{D} \right)\,,
\ee
where ${\cal S}^{(MCS)}$ is the Maxwell Chern-Simons action (\ref{370})
and  ${\cal L}_{D}$ is the free Dirac Lagrangian. Here $G^\mu =\frac{1}{m}\,
\epsilon^{\mu\nu\lambda}\,\partial_{\nu}J_{\lambda}$ is the curl of the fermionic current
introduced in \cite{GMdS}.
Note the  presence of  Thirring like interaction $e^{2}\,J_{\mu}^{2}$ and the change
of the minimal coupling into a magnetic coupling, i.e.,

\be
f_\mu J^\mu \to f_\mu G^\mu \: .
\ee
These terms have appeared naturally as a necessary consequences
of the dualization procedure.  In \cite{GMdS}, they found that this terms are necessary
in order to keep the fermionic sectors of the two models identical.
Extension of this result to bosonic sources (that are field-dependent) should pose no further requirements under our approach \cite{AINRW} but were shown to be troublesome under the master action procedure \cite{GMdS}.


After this digression over the abelian structure of the theory, we are now in position
to study the non-abelian version of the vector self-dual model whose dynamics is given by the action (\ref{380}).
To this end we discuss first in which sense this model possess the self-dual property.
Following the same reasoning as in the abelian case, we define the duality operation as,

\be
{}^{\star}{ F}^{\lambda} \equiv \left[\,\frac{\chi}{m}\,
  \epsilon^{\mu\nu\lambda}\,\left(\partial_{\mu} + { F}_{\mu}\right)
\,\right]\,{ F}_{\nu}\,,
\label{410}
\ee
where the operator inside the square brackets in the right-hand side acts
on the basic field ${ F}_{\nu}$ defining ${}^{\star}{ F}^{\lambda}$ as the dual of ${ F}_{\nu}$. Repeating
this operation, and using the equations of motion obtained by varying  (\ref{380})
with respect to ${ F}_{\lambda}$

\be
{ F}^{\lambda} = \frac{\chi}{2m}\,\epsilon^{\mu\nu\lambda}\,
{ F}_{\mu\nu}\: ,
\label{400}
\ee
we find,
\be
\label{420}
{}^{\star}\left({}^{\star}{ F}_{\mu}\right) = { F}_{\mu}\,,
\ee
thus justifying our terminology and showing the self-dual character of this
model.

Likewise the abelian case, to proceed with the dualization, we begin with the
zeroth-iterated action (\ref{380})
whose variation with respect to ${ F}_{\mu}$ is given by
\be
\label{440}
\delta {\cal S}_\chi = \int\,d^{3}x\,tr\left[\,{ J}^{\mu}\,
                 \delta { F}_{\mu}\,\right]\,,
\ee
with the Noether currents being defined as,
\be
{ J}_{\mu} = -{ F}_{\mu} + \frac{\chi}{2m}\,\epsilon_{\mu\nu\lambda}\,
        { F}^{\mu\nu}\,.
\label{450}
\ee
Our goal is to obtain a non-abelian gauge invariant theory from the above
non-invariant self-dual model. To this end we define the first-iterated action
by a coupling between the currents ${ J}_{\mu}$ and an auxiliary
field ${ B}_{\mu}$,

\be
{\cal S}^{(1)} = {\cal S}_\chi - \,\int\,d^{3}x\,tr\left[\,
        { J}^{\mu}\,{ B}_{\mu}\,\right]\,,
\label{470}
\ee
and whose variation is
\be
\delta {\cal S}^{(1)} = \int\,d^{3}x\,tr\left[\,- \frac{1}{2}\,
\delta \left({ J}^{\mu}\,{ J}_{\mu}\right) -
\delta \left({ J}^{\mu}\,{ B}_{\mu}\right) -
{ J}^{\mu}\,\delta { B}_{\mu}\,\right]\, ,
\label{480}
\ee
where we have used the following transformation rule for the gauging field,

\be
\delta { F}_{\mu} = - \delta { B}_{\mu} - \delta { J}_{\mu}\,.
\label{460}
\ee
This prompt us to define the following second iterated action,

\ba
{\cal S}^{(2)} &=& {\cal S}_\chi + \int\,d^{3}x\,tr\left[\,
\frac{1}{2}\,{ J}^{\mu}\,{ J}_{\mu} -
{ F}^{\mu}\,{ B}_{\mu}\,\right]\nonumber\\
&=& \int\,d^{3}x\, tr\left[\, \frac{\chi}{4m}\,\epsilon^{\mu\nu\lambda}\left({ F}_{\mu\nu}\,
{ F}_{\lambda}
- \frac{2}{3}\,{ F}_{\mu}\,{ F}_{\nu}\,{ F}_{\lambda}
\right) + \frac 12 B_\mu B^\mu\,\right]
\label{490}
\ea
which is gauge invariant after noticing that the transformation rule (\ref{460}) fixes the ${ B}_{\mu}$ field
as
\be
{ B}_{\mu} = -\frac{\chi}{2m}\,\epsilon_{\mu\nu\lambda}\,
{ F}^{\nu\lambda}\,.
\label{500}
\ee
Thanks to the structure of the current (\ref{450}), the action (\ref{490}) can
finally be put in a more familiar presentation as the gauge invariant theory,

\be
\label{1919}
{\cal S}^{(2)}\to {\cal S}^{YMCS}\,,
\ee
the Yang-Mills-Chern-Simons theory
defined in (\ref{510}).  This proves that just as in the abelian case,
the non-abelian self-dual action  defined in (\ref{380}) is physically equivalent to (\ref{510}) for all regimes of the coupling constant.
In the process we have also shown the duality transformation that correctly defines the inherent self-duality property
of action (\ref{380}).

To complete the analysis we discuss the equivalence for the case where dynamical fermions carrying non-abelian charge are coupled to the self-dual bosonic vector fields.  Following the same steps as before we found that the effective action obtained from the second-iterated action, after elimination of the auxiliary field is,

\be
\label{5493}
{\cal S}_{eff} ={\cal S}_{MCS} +  \int d^3 x\: tr\left[{\cal L}_{D} -
\frac { e \chi}{2 m}\epsilon_{\mu\nu\lambda}F^{\mu\nu} J^\lambda + \frac {e^2}{2} J^2\right]
\ee
The result is qualitatively similar to its abelian counterpart.
As in (\ref{9785}) we notice the appearance of a Thirring like current-current interaction in the fermionic sector.  However, the non-minimal magnetic like interaction between the fermionic and bosonic sectors gets modified in this instance due to the presence of a nonlinear piece in the field strength tensor.

In summary, a general method has been developed that establishes dual equivalence between self-dual and topologically massive theories based on the idea of gauge embedding over second-class constrained systems.  The equivalence has been established using an adaptation of the iterative Noether procedure both for abelian and non-abelian self-dual models, including the cases with coupling to dynamical charged fermions.

\section{Application to Non-Abelian Bosonization}

In this section we study the consequences of the non-abelian dual equivalence to bosonization,
the  mapping of a quantum field
theory of interacting fermions onto an equivalent theory of
interacting bosons, in 2 + 1 dimensions. Bosonization
was developed in the context of the two-dimensional scalar field theory and has been one
of the main tools available to investigate the non-perturbative behavior of some interactive field theories \cite{2dbos}.
For some time this concept was thought to be an exclusive property of two-dimensional space-times where spin is absent and one cannot distinguish between bosons and fermions.
It was only recently that this powerful technique were extended to higher dimensional space-times \cite{BLQ,FGM}\cite{FS,B}.
The bosonization mapping in D=3, first discussed by Polyakov\cite{P}, shows that this is a relevant issue in the context of transmutation of spin and statistics in three dimensions.
The equivalence of the three dimensional effective electromagnetic action of the $CP^1$ model with a charged massive fermion to lowest order in inverse (fermion) mass has been proposed by
Deser and Redlich \cite{DR}.  Using their results bosonization was extended to three dimensions in the 1/m expansion \cite{FS}. These endeavor has led to promising results in diverse areas such as, for instance, the understanding of the universal behavior of the Hall conductance in interactive fermion systems \cite{BOS}.

For higher dimensions, due to the absence of an operator mapping
a la Mandelstan, the situation is more complicate and even the bosonization identities extracted from these procedures relating the
fermionic current with the bosonic topological current is a consequence of a
non-trivial current algebra.  Moreover, contrary to the two dimensional case, in dimensions higher
than two there are no exact results with the exception of the current mapping \cite{LMNS,LG}.
Besides, while the two-dimensional fermionic determinant can be exactly computed, here it is neither exact nor complete, having a non-local structure.
However, for the large mass limit in the one-loop of perturbative evaluation, a local expression materializes.
This procedure, is in a sense, opposite to what is done in
1 + 1 dimensions where bosonization is a set of operator identities
valid at length scales short compared with the Compton
wavelength of the fermions while in D=3 only the
long distance regime is considered.

In this section we review how the low energy sector
of a theory of massive self-interacting, G-charged fermions, the massive Thirring Model
in 2 + 1 dimensions, can be bosonized  into a gauge theory, the Yang-Mills-Chern-Simons
gauge theory thanks to the results of the preceeding section.

\subsection{The Non-Abelian Mapping}

In the sequence we investigate the problem of  identifying a bosonic
equivalent of a three dimensional theory of self-interacting fermions with symmetry group G and show how it is possible to bosonize the low-energy regime of the theory.
We follow the same strategy as in reference \cite{FS} but follow the notation of \cite{BFMS} that is slightly different than \cite{FS}. We seek a bosonic theory which reproduces correctly the low-energy regime of the massive fermionic theory.  To begin with we define the $G$-current,

\be
\label{NB10}
j^{a\mu} = \bar\psi^i t^a_{ij} \gamma^{\mu} \psi^j\, ,
\ee
where $\psi^i$ are N two-component Dirac spinors in the fundamental
representation of $G$, $i,j = 1,\ldots , N$ and $a= 1,\ldots ,\mbox{dim}\: G$.
Here $t^a$ and $f^{abc}$ are the generators  and the structure constants of the symmetry group $G$, respectively and $j^{a\mu}$ is a $G$-current.

The (Euclidean) fermionic partition function for the three-dimensional massive
Thirring model is,

\be
\label{BB10}
{\cal Z}_{Th} = \int \,{\cal{D}}\bar\psi {\cal{D}}\psi\; e^{-\int \left ( \bar\psi^i (\dslash + m) \psi^i -\frac{g^2}{2N} j^{a\mu}j_{\mu}^a
\right ) d^3x}
\ee
with the coupling constant $g^2$ having dimensions of inverse mass.
Next we eliminate the quartic interaction through a Legendre transformation

\be
e^{\frac{g^2}{2} \int\!d^3\!x\, j^{a\mu}j_{\mu}^a } = \int{\cal D} a_{\mu}
e^{-\int\!d^3\!x\, tr(\frac{1}{2g^2}a^{\mu}a_{\mu} + j^{\mu}a_{\mu})}
\label{NB20}
\ee
up to a multiplicative normalization constant, where we have introduced a vector field
$a_{\mu}$ taking values in the Lie algebra of G.  After integration of the fermionic degrees of freedom, the partition function reduces to,

\be
{\cal Z}_{Th} = \det (i\dslash + m + \aslash) \;\int {\cal{D}} a_{\mu}
e^{-\frac{1}{2g^2}\!\int d^3 x\:  tr \left(a^{\mu}a_{\mu}\right)}  .
\label{NB30}
\ee

The determinant of the Dirac operator is an unbounded operator and requires regularization.
For D=2 this determinant can be computed exactly, both for abelian and non-abelian symmetries.
Based on general grounds only, one may say that this determinant consists of a Chern-Simons action standing as the leading term plus an infinite series of terms depending on the dual of the vector field, $\tilde F_\mu \sim \epsilon_{\mu\nu\lambda}\partial^\nu A^\lambda$, including those terms that are non-local and non-quadratic in $\tilde F_\mu$.
For the D=3 the actual computation of this determinant will give parity breaking and parity conserving
terms that are computed in powers of the inverse mass,
\be
\ln \det (i\dslash + m + \aslash) =   \frac{\chi}{16\pi} {\cal S}_{CS}[a]
+  I_{PC}[a] + O(\partial^2/m^2)  \: .
\label{NB40}
\ee
Here ${\cal S}_{CS}$ given by

\be
\label{NB50}
{\cal S}_{CS}[a] = \int d^3 x\: i \epsilon^{\mu\nu\lambda}\, tr
(f_{\mu \nu} a_{\lambda} - \frac{2}{3} a_{\mu}a_{\nu}a_{\lambda})  ,
\ee
is the non-abelian Chern-Simons action and the parity conserving contributions, in first-order, is the Yang-Mills action

\be
I_{PC}[a] = - \frac{1}{24\pi  m}\, tr\int d^3 x\: f^{\mu\nu} f_{\mu\nu}  ,
\label{NB60}
\ee
where

\be
f_{\mu\nu} = \partial_{\mu}a_{\nu} - \partial_{\nu}a_{\mu} + [a_{\mu},a_{\nu}]  .
\label{NB70}
\ee

In the low energy regime, only the Chern-Simons action survives yielding a closed expression for the determinant as,

\be
\lim_{m\to\infty} \det (i\dslash + m + \aslash) =   \frac{\chi}{16\pi} {\cal S}_{CS}[a]
\label{NB75}
\ee
Using this result we can write ${\cal Z}_{Th}$ in the form

\be
{\cal Z}_{Th} = \int {\cal{D}} a_{\mu}  \exp(-{\cal S}_{SD}[a])  ,
\label{NB80}
\ee
where ${\cal S}_{SD}$ is the non-abelian version of the self-dual action  introduced in \cite{TPvN},

\be
{\cal S}_{SD}[a] = \frac 1{2g^2}\int d^3 x\: tr \left(a_\mu a^\mu \right) - \frac \chi{16\pi} {\cal S}_{CS}[a]
\ee
Therefore, to leading order in $1/m$ we have established the identification ${\cal Z}_{Th} \approx  {\cal Z}_{SD} $.


Now, recalling that the model with dynamics defined by the non-abelian self-dual action is equivalent to
the Yang-Mills-Chern-Simons theory, we use this connection to establish the equivalence of the non-abelian massive Thirring
model and the YMCS theory as

\be
{\cal Z}_{Th} \approx  {\cal Z}_{YMCS} \: .
\label{NB100}
\ee
It is interesting to observe that the Thirring coupling constant
$g^2/N$ in the fermionic model is mapped into the inverse mass spin $1$ massive excitation,
$m=\pi/g^2$.

Now comes an important observation.  Unlike the master approach, our result is valid for all values of the coupling constant.  The proof, based on the use of an ``interpolating Action'' ${\cal S}_I$, is seen to run into trouble in the non-abelian case. That the non-abelian extension of this kind of equivalences is more
involved was already recognized in \cite{DJ} and \cite{BFMS}, and shown that the non-abelian self-dual
action {\it is not}
equivalent to a Yang-Mills-Chern-Simons theory (the natural extension
of the abelian MCS theory) but to a model where the
Yang-Mills term vanishes in the limit $g^2 \to 0$ \cite{BFMS}.

\subsection{Current Identities}

To infer the bosonization identities for the currents which derive from the equivalence
found in the last section, we add a source for the Thirring current leading to the following
functional generator,

\ba
{\cal Z}_{Th}[b] &=& \int {\cal{D}}\bar\psi {\cal{D}}\psi {\cal{D}} a_\mu
e^{-\int d^3 x\: \left[\bar\psi (i\dslash + m + \aslash + \bslash)\psi
+\frac{1}{2g^2}\, tr\left( a^\mu a_\mu\right)\right]}\nonumber\\
&=&  e^{-\frac{1}{2g^2} \, tr \int d^3 x\: b_\mu b^\mu} \cdot
\int {\cal{D}} a_\mu e^{-{\cal S}_{SD}[a] + \frac{1}{g^2}\, tr \int d^3 x\: b_\mu a^\mu}
\label{CC10}
\ea
after shifting $a_\mu \to a_\mu - b_\mu$.

In order to connect this with the Yang-Mills-Chern-Simons system
we repeat the steps of the last section to obtain,

\be
{\cal Z}_{Th}[b_{\mu}] \approx {\cal Z}_{MCS}[b_{\mu}]
\label{CC30}
\ee


We have therefore established, to order $1/m$, the connection between the Thirring and self-dual models in the non-abelian context, now in the presence of sources.
This is, in its most general form, the result we were after.
It provides a complete low-energy bosonization prescription,
valid for any $g^2$, of the matrix elements of the fermionic
current.
From (\ref{CC30}) we see, from simple differentiation w.r.t.
the source, that the bosonization
rule for the fermion current, to leading order in $1/m$, reads

\be
j_\mu^a \to  i \frac{\chi}{8\pi} \epsilon^{\mu \nu\lambda}
f_{\nu\lambda}^a
\label{CC40}
\ee
Eq.(\ref{CC40}) gives a natural non-abelian extension of the
abelian bosonization rule obtained in \cite{FS}, extended to the weak coupling
non-abelian case in \cite{BFMS}. In contrast to these results,  the bosonization identity (\ref{CC40}), was shown to be an exact recipe in \cite{LMNS,LG}.

In the
abelian case one can interpret the $(2+1)$ dimensional bosonization
formula as the analog of the
$(1+1)$-dimensional result $\bar\psi\gamma^{\mu}\psi \to
({i}/{\sqrt{\pi}}) \epsilon_{\mu\nu}\partial^\nu\phi$ while in this case it should be considered as the analog of the Wess-Zumino-Witten currents.
Notice that as in the abelian case the bosonized expression for the fermion current
is topologically conserved.

We thus see that
the non-abelian bosonization of free, G-charged massive fermions in
$2+1$ dimensions leads to the non-abelian Chern-Simons
theory, with the fermionic current being mapped to the dual of the
gauge field strength.  This result holds only
for length scales large compared with the Compton wavelength of the
fermion, since our results were obtained for large fermion mass.
It is important to notice that the limit $g^2 \to 0$ used in earlier
approaches corresponding to free fermions (but not to an
abelian gauge theory) was not taken here at any stage.
This is important since Yang-Mills coupling is proportional to $g^2 $, which
is why we are left with a Yang-Mills-Chern-Simons action and not the
pure Chern-Simons theory of \cite{BFMS}.

\section{Conclusions}

The rationale of different phenomena in planar physics have greatly benefitted from the use of 2+1 dimensional
field theories including the parity breaking Chern-Simons term.
In this scenario it is important to establish connections among different models so that a
unifying picture emerges.  In this context we have shown, in earlier work, that the
soldering formalism has established a direct link between self-dual models of opposite helicities
with the Proca model \cite{BW}.  Other instances includes the recent extension of the functional bosonization program
interpolating from fermions to bosons in a coherent picture \cite{FS,BFMS}.

In the context of the first it has been argued that the soldering formalism is equivalent
to canonical transformation albeit in the Lagrangian side while for the later
the mentioned mapping between SD and MCS models has been used to establish a formal equivalence
between the partition functions of the abelian version of MTM and a theory of interacting bosons.
The non-abelian extension of this analysis, for the full range of the coupling constant, has been the main concern of the present work since only partial results were reported in the literature.  Other directions
have also been investigated, with new results, that includes the proof of the self-duality property of (\ref{380}) and the coupling with G-charged dynamical matter fields.

Our analysis has shown how the gauge lifting approach sheds light on the question of
dual equivalence between SD and topologically massive theories with new results for the non-abelian case.  This discussion becomes the central issue when deriving bosonization
rules in D=3, for fermions carrying non-abelian charges since, up to date, only prescriptions based on
the Master action of Ref.\cite{DJ} were used, apart from \cite{BBG} with conclusions consistent
with \cite{BFMS}.
These derivations of the bosonization mappings
suffered from well known difficulties related to the dual equivalence, restricting the results to be limited to weak
coupling constant only. Therefore, regarding the non-abelian bosonization in D=3 dimensions, we believe that the method developed here, which is simpler and
better suited to deal with non-abelian symmetries, completes the program initiated in \cite{BFMS} and confirms the exact identities found in \cite{LMNS,LG}. This new approach has also been used with dynamical fermionic fields leading naturally to the necessity
of a Thirring like term to establish the equivalence of the fermionic sectors in both sides.
Such equivalence may be extended to the scalar case \cite{AINRW}.

The bosonization for $D\geq 4$ poses no difficulties as long as the
fermionic determinant can be evaluated in some approximation and is expected to yield a
gauge invariant piece. This is of importance since the description of charged fermionic fields in terms of gauge fields has brought new perspectives and a deeper insight on the non perturbative dynamics of planar physics \cite{elcio} that might be extended to higher dimensions.

\end{document}